\documentstyle[prd,aps,eqsecnum,preprint,tighten,psfig,floats]{revtex}
\begin{document}
\draft
\preprint{FSU-HEP-960313}

\title{Fully differential heavy quark contributions to the photon \\ 
      structure functions $F_{2}^{\gamma}(x,Q^2)$ 
      and $F_{L}^{\gamma}(x,Q^2)$ \\
      at next-to-leading order QCD}

\author{B. W. Harris and J. F. Owens}

\address{Physics Department \\ Florida State University \\
Tallahassee, Florida 32306-3016, USA}

\date{March 1996}

\maketitle

\begin{abstract}
Fully differential heavy quark contributions to the photon structure 
functions in deeply inelastic scattering are computed in next-to-leading 
order QCD, including both the direct and resolved contributions.  
A variety of distributions are presented and discussed.  Several of the 
distributions show marked differences between the resolved and direct 
cases due primarily to the presense of the gluon distribution in the 
former and lack thereof in the later.
\end{abstract}


%---------------Section 1--------------------------
\section{Introduction}
%--------------------------------------------------
Hard scattering processes in two photon collisions have long been recognized 
as playing an important role in the investigation of QCD. Examples include the 
photon structure function, the production of high-p$_T$ jets, 
and heavy quark production, to name just a few. Such processes can yield
information pertaining to the hard-scattering parton-level cross sections, the
distributions of partons in one or both initial photons, and the fragmentation
of partons into hadrons in the final state. The information obtained thereby is
complementary to that obtained in other hard scattering processes. New 
developments in these and related areas have recently been reviewed in 
\cite{work}. 

Activity in the area of two photon physics has increased recently with new data
becoming available from several groups and with the 
initiation of the LEP2 research program. In particular, it is anticipated that
substantial samples of charm quark production data will soon be obtained. This
data sample can be divided into three subsets, each of which is treated
differently from a theoretical standpoint. First, both photons which originate
from the $e^+$ and $e^-$ beams can be nearly on-shell. This is by far the
largest subset and the relevant theoretical issues have been discussed in
\cite{review}, for example. The second largest subset, and the one addressed
in this paper, consists of one photon being nearly real and the other being
substantially off-shell, with a negative squared invariant mass satisfying 
$Q^2>2\;$(GeV/c)$^2$, for example. The third subset consists of both photons 
being far off-shell, but the rate is substantially reduced and this case will 
not be addressed here.

When an on-shell photon interacts in a hard-scattering process, two cases 
must be considered.  First, the photon may interact via the pointlike 
electromagnetic coupling to quarks.  This component is often referred to 
in the literature as the direct or pointlike piece.  In the case considered 
here, this contribution is calculable in perturbation theory once the heavy 
quark mass and the strong coupling are specified.  
Second, the photon may undergo a fluctuation into a hadronic state. 
The connection to the hard scattering is then made through parton 
distributions in the photon. These can be obtained,
for example, from studies of the photon structure function or the production of
high-p$_T$ jets in two photon collisions or in photoproduction. On the other
hand, the sensitivity of heavy quark production to these distributions may be
used to provide additional constraints on them.

The purpose of this work is to present the fully differential heavy 
quark contributions to the photon structure functions. 
These may be compared directly with experimentally measured structure 
functions, should they become available, or embedded in a larger 
Monte Carlo program for generation of heavy hadron events.   
Heavy quark correlations have also been studied theoretically 
in hadroproduction \cite{mnr}, photoproduction \cite{fmnr2}, 
electroproduction \cite{hs}, and recently in direct photon-photon 
collisions \cite{kl} as well as experimentally \cite{hvqexp}.  

Our calculation is performed using the subtraction method which is 
based on the replacement of divergent (soft or collinear) terms in 
the squared matrix elements by generalized distributions.  
The method was first used in the context of electron-positron 
annihilation \cite{ellis} and its essence is nicely 
described and compared to the phase-space slicing 
method \cite{owens} in the introduction of a paper by Kunszt and Soper 
\cite{ks}. 

In Sec.\ II the method of calculation is reviewed and analytical 
results presented.  Numerical results and a discussion of related physics
issues are presented in Sec.\ III and some conclusions are given in 
Sec.\ IV.

%---------------Section 2--------------------------
\section{Method}
%--------------------------------------------------
Deep inelastic photon structure functions $F_{k}^{\gamma}(x,Q^2)$ $(k=2,L)$ 
are measured in $e^+ e^-$ collisions via the process
\begin{equation}
\label{ee}
e^-(l) + e^+ \rightarrow e^-(l^{\prime}) + e^+ + X
\end{equation}
where $X$ denotes any hadronic final state allowed by quantum number 
conservation.  See Figure 1.  
When the outgoing electron is tagged the reaction is dominated by
\begin{equation}
\gamma^{\ast}(q) + \gamma(p) \rightarrow X
\end{equation}
where one of the photons is highly virtual and the other is near 
mass shell.  For heavy quark production the relevant reaction is 
\begin{equation}
\gamma^{\ast}(q) + \gamma(p) \rightarrow Q(p_1) + \bar{Q}(p_2) + X 
\end{equation}
and the cross section is given by
\begin{equation}
\frac{d^2\sigma(e^+ e^- \rightarrow Q \bar{Q} e^+ e^- X)}{dxdy} 
= \frac{2\pi\alpha^2}{Q^4} S \left[ \left\{ 1 + (1-y)^2 \right\} 
F_2^{\gamma}(x,Q^2,m^2) - y^2 F_L^{\gamma}(x,Q^2,m^2) \right]
\end{equation}
after integration over the azimuthal angle between the plane containing 
the incoming and outgoing electron and the plane containing $\gamma(p)$ and 
the outgoing heavy quark.
Here $q=l-l^{\prime}$, $Q^2=-q^2$, $x=Q^2/2p \cdot q$, and 
$y=p \cdot q / p \cdot l$.  
The heavy quark contributions to the photon structure functions are
denoted by $F_k^{\gamma}(x,Q^2,m^2)\;$ $(k=2,L)$ 
with the heavy quark mass $m$ shown explicitly.  
The electromagnetic coupling is $\alpha = e^2 / 4 \pi$ and 
$S$ is the square of the c.\ m.\ energy of the electron-positron system.

We follow the terminology of \cite{dg} in distinguishing 
between direct and resolved photon contributions.  Direct 
contributions are those reactions where the incident photon 
participates directly in the hard scattering process.  Resolved 
contributions are those reactions where the incident photon 
participates in the hard scattering via one of its constituents.
For heavy quark production at next-to-leading order, the direct 
component is independent of the parton distributions in the photon.  
This is expressed as 
\begin{equation}
F^{\gamma}_{k}(x,Q^2,m^2,\mu_f^2) = F^{\gamma,R}_{k}(x,Q^2,m^2,\mu_f^2) + 
                                  F^{\gamma,D}_{k}(x,Q^2,m^2).
\end{equation}
The mass factorization scale $\mu_f$ is explicitly shown.  
It is absent in the second term because any potential collinear 
singularities are regulated by the heavy quark mass $m$.    
Calculation of $F^{\gamma,D}_{k}$ and $F^{\gamma,R}_{k}$ in fully 
differential form at NLO will be discussed in the following sections and 
represents a generalization of the fully and single inclusive results 
already available in \cite{LRSvN}.

In our computation we use the subtraction method which is
based on the replacement of divergent (soft or collinear) terms in
the squared matrix elements by generalized distributions.
This allows us to isolate the soft and collinear poles within the
framework of dimensional regularization without calculating all the
phase space integrals in a space-time dimension $n \ne 4$ as is usually
required in a traditional single particle inclusive computation.
In this method the expressions for the squared matrix elements in the
collinear limit appear in a factorized form, where poles in $n-4$ multiply
splitting functions and lower order squared matrix elements.
The cancellation of collinear singularities is then performed using 
mass factorization \cite{CSS}.   The expressions for the squared matrix
elements in the soft limit also appear in a factorized form where poles in 
$n-4$ multiply lower order squared matrix elements.  The cancellation of soft
singularities takes place upon adding the contributions from the renormalized
virtual diagrams.  Since the final result is in four-dimensional space time,
we can compute all relevant phase space integrations using standard Monte
Carlo integration techniques \cite{lepage} and produce histograms for
exclusive, semi-inclusive, or inclusive quantities related to any of the
outgoing particles.  We can also incorporate various
experimental cuts defined in terms of partonic variables.

\subsection{Direct component}

To order $\alpha_s$, the subprocesses where the incident on-shell photon 
participates directly in the hard scattering process are given by 
\begin{equation}
\label{d22}
\gamma^{\ast} (q) \; \gamma (k_1) \rightarrow Q(p_1) \; \bar{Q}(p_2) 
\end{equation}
\begin{equation}
\label{d23}
\gamma^{\ast} (q) \; \gamma (k_1) \rightarrow Q(p_1) \; \bar{Q}(p_2) \; g(k_2).
\end{equation}
For future reference we introduce the invariants $s=(q+k_1)^2$ and  
$s_5=(p_1+p_2)^2$ and define additionally, $s^{\prime}=s-q^2$ and 
$s_5^{\prime}=s_5-q^2$.
The direct piece of the fully differential heavy quark contribution 
to the photon structure function follows from 
\begin{equation}
d F^{\gamma,D}_k (x,Q^2,m^2) = \frac{Q^2}{4 \pi^2 \alpha} \; d \sigma_k
\end{equation}
with $k=2,L$ and $d\sigma_2 = d\sigma_G + 3 d\sigma_L / 2$.  The 
virtual photon-photon cross sections are summarized as 
\begin{equation}
\label{sig1}
d \sigma_i = d \sigma^{(0)}_i + d \sigma^{(sv)}_i + d \sigma^{(f)}_i
\end{equation}
for $i=G,L$.  The first term is the lowest order contribution 
from (\ref{d22}) and is given by 
\begin{equation}
d \sigma^{(0)}_i = C_i \; M_i^{(0)} \; d \Gamma_2
\end{equation}
where $C_G = 1/(8 s^{\prime})$ and $C_L=1/(4 s^{\prime})$. 
$d \Gamma_2$ is the two body differential phase space factor.  
$M_i^{(0)} = 128 \pi^2 e_H^4 \alpha^2 N B_{i,QED}$ where 
$e_H$ is the charge of the heavy quark in units of $e$, $N=3$ for 
$SU(3)$, and $B_{i,QED}$ may be found in \cite{hs}, for example.

The second term in (\ref{sig1}) is the finite contribution remaining after the 
addition of the soft contribution from (\ref{d23}) and the interference of 
(\ref{d22}) with its renormalized one loop corrections.  
It is expressible as 
\begin{equation}
d \sigma^{(sv)}_i = C_i \; M_i^{(sv)} \; d \Gamma_2
\end{equation}
where $M_i^{(sv)} = 4 \pi \alpha_s e_H^4 \alpha^2 N C_F V_{i,QED}$.  
$\alpha_s$ is the strong coupling and $C_F=4/3$ for $SU(3)$.  
The expression for $V_{i,QED}$ is too long to be presented here but 
may be obtained from the authors upon request.
The third term is a finite contribution given by 
\begin{equation}
\label{finite}
d \sigma^{(f)}_i = 2 \left( \frac{1}{16 \pi^2} \right)^2 
C_i \beta_5 \frac{s}{(s^{\prime})^2} \left( \frac{1}{1-z} 
\right)_{\tilde{\rho}} \frac{f_i(z,y,\theta_1,\theta_2)}{(1+y)(1-y)} 
dz dy \sin \theta_1 d \theta_1 d \theta_2.
\end{equation}
Here we have defined $\beta_5=\sqrt{1-4m^2/s_5}$, 
$z=s^{\prime}_5/s^{\prime}$, and $y$ denotes the cosine of the angle 
between ${\bf q}$ and ${\bf k}_2$ 
in the $\gamma^{\ast} \gamma$ center of mass frame.
$f_i(z,y,\theta_1,\theta_2)$ is the coefficient of $C_F^2$ in 
$f_i^g(z,y,\theta_1,\theta_2)$ given in \cite{hs}; it is, up to 
overall factors, the squared matrix element for the 
subprocess in Eq.\ (\ref{d23}) times $(1-z)^2(1-y)(1+y)$.  
Note that $f_i$ is finite in the soft $(k_2 \rightarrow 0)$ limit.  
The $\tilde{\rho}$ distribution is defined by 
\begin{equation}
\label{rho}
\int_{\tilde{\rho}}^1 dz f(z) \left( \frac{1}{1-z} \right)_{\tilde{\rho}} = 
\int_{\tilde{\rho}}^1 dz \frac{f(z)-f(1)}{1-z}.
\end{equation}
The lower limit of the RHS of (\ref{rho}), when inserted in (\ref{finite}), 
cancels exactly against a corresponding piece in $V_{i,QED}$ so that the 
overall expression is independent of $\tilde{\rho}$.

\subsection{Resolved component}

To order $\alpha_s^2$, the subprocesses where the incident on-shell photon 
participates in the hard scattering via one of its constituents are 
\begin{eqnarray}
\gamma^{\ast} \; g & \rightarrow & Q \; \bar{Q} \nonumber \\
\gamma^{\ast} \; g & \rightarrow & Q \; \bar{Q} \; g \nonumber \\
\gamma^{\ast} \; q & \rightarrow & Q \; \bar{Q} \; q \nonumber \\
\gamma^{\ast} \; \bar{q} & \rightarrow & Q \; \bar{Q} \; \bar{q} \; .
\end{eqnarray}
The calculation of the fully differential heavy quark contributions to the 
photon structure function proceeds along the same lines as those for 
the ${\em proton}$ structure functions and will not be repeated here.  
The interested reader is invited to consult \cite{hs} for full details.  
The final result is of the form 
\begin{eqnarray}
\label{fhad}
F_{k}^{\gamma,R}(x,Q^2,m^2,\mu^2) &=& \frac{Q^2 \alpha_s(\mu^2)}{4\pi^2 m^2} 
\int_{\xi_{\rm min}}^1 \frac{d\xi}{\xi} \left\{ \left[ c^{(0)}_{k,g} + 
4 \pi \alpha_s(\mu^2) \left( c^{(1)}_{k,g} 
+ \bar c^{(1)}_{k,g} \ln \frac{\mu^2}{m^2} \right)
\right] e_H^2 f_{g/\gamma}(\xi,\mu^2) \right. \nonumber \\ 
&+& \left. 4 \pi \alpha_s(\mu^2) \sum_{i=q,\bar q} f_{i/\gamma}(\xi,\mu^2) 
\left[ e_H^2 \left( c^{(1)}_{k,i} + \bar c^{(1)}_{k,i} \ln \frac{\mu^2}{m^2} 
\right) + e^2_i \, d^{(1)}_{k,i} + e_i\, e_H \, o^{(1)}_{k,i} \, 
\right] \right\} \nonumber \\ &&
\end{eqnarray}
where $k = 2,L$.  The lower boundary on the integration is
$\xi_{\rm min} = x(4m^2+Q^2)/Q^2$. 
The parton momentum distributions in the photon are denoted by 
$f_{i/\gamma}(\xi,\mu^2)$.  The mass factorization scale $\mu_f$ has been set 
equal to the renormalization scale $\mu_r$ and is denoted by $\mu$.
Finally,  $c^{(l)}_{k,i}$ and $\overline c^{(1)}_{k,i}\,,
(l=0,1)$,  and $d^{(1)}_{k,i}$ and 
$o^{(1)}_{k,i}$ are scale independent parton coefficient functions. 
In Eq.\ (\ref{fhad}) the coefficient functions are distinguished 
by their origin: $c^{(l)}_{k,i}$ and $ \overline c^{(1)}_{k,i}$ 
originate from the virtual photon-heavy quark coupling and 
therefore appear for both 
charged and neutral parton-induced reactions; $d^{(1)}_{k,i}$ arise 
from the virtual photon-light quark coupling; $o^{(1)}_{k,i}$ come 
from the interference between these processes.  
All charges are in units of $e$.  
We have isolated the scale dependent terms, 
proportional to $\overline c^{(1)}_{k,i}$.
Finally, note that Eq.\ (\ref{fhad}) only holds for $Q^2>0$. 
When $Q^2=0$ one encounters double resolved processes which are not 
discussed here.

%---------------Section 3--------------------------
\section{Results}
%--------------------------------------------------
Using the results of the previous section we have constructed a program for 
the fully differential heavy quark contributions to the photon structure 
functions.  With this program we can calculate fully inclusive, single 
inclusive, and fully differential structure functions.  
The program uses Monte Carlo integration so it is possible to implement 
experimental cuts, provided that they are defined in terms of partonic 
variables.
Additional integrations and final state fragmentation may be added to 
produce hadronic cross sections.
We have verified, to better that $1\%$, that our program produces fully 
inclusive structure functions that match those already available \cite{check}.

We use a mass of $1.5\;$GeV for the charm contribution to the photon 
structure functions and neglect the bottom quark content. 
The latter contribution is suppressed by the larger mass and smaller charge 
of the $b-$quark.  The strong coupling $\alpha_s$ evaluated with $N_f=3$ at 
one loop accuracy is used in the LO results.  
For the NLO results the two loop value is used. 
The GRV photon parton distribution functions \cite{grv} in the 
$\overline{{\rm MS}}$ scheme are used.  This parameterization has 
$\Lambda_{LO}^{(N_f=3)}=232\;$MeV and $\Lambda_{HO}^{(N_f=3)}=248\;$MeV 
for the leading order and higher order parameterizations, respectively.

In Fig.\ 2 the direct and resolved contributions to 
$F_2^{\gamma}(x,Q^2,m_c^2)/\alpha$ at $Q^2=10\;$(GeV/c)$^2$ 
are shown as functions of $x$.  
We choose the renormalization and factorization scales to be equal to 
$\mu_0$ where $\mu_0^2=Q^2+4m^2_c$.  This choice reduces to 
$\mu^2=Q^2$ for the electroproduction of massless quarks
and $\mu^2=4m^2_c$ for the photoproduction of charm quarks.  
The dashed lines are the LO contributions and the solid lines are the 
NLO contributions.
Clearly, the two components dominate in different $x$ regions, as was 
noted previously in \cite{LRSvN}.  
The direct component peaks just above threshold while the resolved 
component peaks at small $x$ due to a sharp rise in the gluon density 
in the photon.

Typically, the photon-quark initiated process contribution to 
$F_2^{\gamma}(x,Q^2,m_c^2)$ is so small that it may be neglected.
However, this is a scale dependent statement which must be verified.  
Using our results we have checked that formally the variation of 
$F_2^{\gamma}(x,Q^2,m_c^2,\mu_r^2,\mu_f^2)$ with respect 
to $\mu_r$ or $\mu_f$ is one higher power in $\alpha_s$.  
We have also studied this numerically by keeping $\mu_r$ and $\mu_f$ separate 
and distinct and examining the change in the result under their variations.  
At $x=10^{-3}$, when $\mu_f$ is varied between $2\mu_0$ and $\mu_0/2$,  
$F_2^{\gamma}(x,Q^2,m_c^2,\mu_r^2=\mu_0^2,\mu_f^2)$ changes by $+6\%$ and 
$-14\%$, respectively, at NLO to be compared with $+26\%$ and $-29\%$ at LO.  
Under the same variation, the sum of quark and anti-quark initiated process 
contributions changes from $-10\%$ to $+5\%$.  These contributions are 
around $-1\%$ at the central value.  
Also at $x=10^{-3}$ when $\mu_r$ is varied between $2 \mu_0$ and
$\mu_0/2$, $F_2^{\gamma}(x,Q^2,m_c^2,\mu_r^2,\mu_f^2=\mu_0^2)$ changes
by $-11\%$ and $+13\%$, respectively, at NLO to be compared with $-19\%$ and
$+31\%$ at LO.  Under the same variation, the sum of quark and
anti-quark initiated process contributions changes from $-1\%$ to $-2\%$.
We therefore conclude that at small $x$ the dominance of the resolved 
over the direct contribution to $F_2^{\gamma}(x,Q^2,m_c^2)$ could 
provide information on the gluon content of the photon with no need to 
consider the (anti)quark initiated processes.

Let us next examine a sample of the various differential structure functions 
that can be produced with our results.  For these plots we have not applied 
any cuts.  We choose the renormalization and factorization scales to be 
equal to $\mu_0$ with $\mu_0^2=Q^2+4m^2_c+m_x^2$ where 
$m_x^2=(P_t^2(c)+P_t^2(\bar{c}))/2$.
Note that there are many possible choices of scale.  We have simply 
chosen to add in the average of the squares of the transverse momenta of 
the heavy quarks.  For illustration purposes, we consider the limiting 
cases of small and large $x$ in the following figures.  In all plots we
take the $z$ axis to be in the direction of the $\gamma^{\ast}$ in the
$\gamma^{\ast} \gamma$ c.\ m.\ frame.

We begin by examining the invariant mass distribution of the charm-anticharm 
pair.  The results for 
$d (F_2^{\gamma}(x,Q^2,m_c^2)/ \alpha) / d M_{c \bar{c}}$
at $Q^2=10\;$(GeV/c)$^2$ are shown in Fig.\ 3 with the LO results shown as 
dashed lines and the NLO results shown as solid lines.  At $x=10^{-3}$ 
the resolved contribution dominates whereas at $x=0.3$ the direct contribution 
dominates.  For the resolved case, the NLO curve is flatter than the LO curve  
because the quark and anti-quark initiated processes reach their largest 
negative value just above threshold.  This compensates for the gluon 
initiated process which tends to make the curve steeper due to negative 
contributions at large invariant mass.  For the direct 
case, at leading order, the invariant mass is fixed at $\sqrt{s}$ with 
$s=Q^2(1-x)/x$.  
At NLO, the radiated gluon prefers to be soft 
so the correction rises from zero at threshold $2m_c$ to 
a maximum at $\sqrt{Q^2(1-x)/x}$.

The transverse momentum distributions 
$d (F_2^{\gamma}(x,Q^2,m_c^2)/ \alpha) / d P_t(c)$ at
$Q^2=10\;$(GeV/c)$^2$ are shown in Fig.\ 4.  $P_t(c)$ is the transverse 
momentum of the charm quark relative to the $z$ axis defined above. 
The LO results are shown as 
dashed lines and the NLO results are shown as solid lines.  The results for 
$x=10^{-3}$ are shown in Fig.\ 4(a) and for $x=0.3$ in Fig.\ 4(b).  Also shown 
for comparison in the inset is the LO result for 
$Q^2=100\;$(GeV/c)$^2$ and $x=0.01$.  
For the direct case there is a Jacobian peak at 
$P_t(c)^{\rm max} = \sqrt{s/4-m_c^2}$ and a factor of $P_t(c)$ which forces 
a zero at $P_t(c)=0$.  Depending on $\sqrt{s}$ a peak may develop 
also at small but non-zero $P_t(c)$ as shown in the inset.  However, at 
these values of $x$ and $Q^2$ the resolved contribution dominates.

At leading order $d (F_2^{\gamma}(x,Q^2,m_c^2)/ \alpha) / d P_t(c \bar{c})$, 
shown in Fig.\ 5 at $Q^2=10\;$(GeV/c)$^2$, is a delta function 
at $P_t(c \bar{c})=0$ for both the direct and resolved cases.  
$P_t(c \bar{c})$ is the sum of the charm and anticharm transverse 
momenta and is identically zero at LO but is balanced by the transverse 
momentum of the radiated gluon at NLO.  The resolved contribution has large 
negative contributions at $P_t(c \bar{c})=0$ due to a combination of soft and 
collinear subtractions.  This accounts for the reduced value in the 
first bin.  If fact, for very narrow binning these distributions 
become negative at $P_t(c \bar{c})=0$.
The direct contribution has only soft subtractions which 
are much less extreme.

Another distribution that shows similar effects is 
$d (F_2^{\gamma}(x,Q^2,m_c^2)/ \alpha) / d (\Delta \phi)$ 
shown in Fig.\ 6 at $Q^2=10\;$(GeV/c)$^2$.  $\Delta \phi$ is the 
azimuthal angle between the charm and anti-charm in the plane 
perpendicular to the $z$ axis.  
At LO, for both the resolved and direct cases, this distribution is a 
delta function at the back-to-back configuration $\Delta \phi = \pi$.
Again there is a depletion at NLO in the bin nearest $\Delta \phi = \pi$.   
This is most noticeable in the resolved contribution where there 
are both collinear and soft subtractions.

The role of parton distributions in the photon can be further seen by 
examining various (pseudo)rapidity dis\-tri\-butions.
$d (F_2^{\gamma}(x,Q^2,m_c^2)/ \alpha) / d y (c)$ is shown in Fig.\ 7 and 
$d (F_2^{\gamma}(x,Q^2,m_c^2)/ \alpha) / d \eta (c)$ is shown in Fig.\ 8, 
both at $Q^2=10\;$(GeV/c)$^2$.  The charm quark rapidity $y(c)$ is defined 
by $y=1/2 \ln [(E+p_z)/(E-p_z)]$ and the pseudo-rapidity $\eta(c)$ 
is defined by $\eta(c)=1/2 \ln [(1+\cos \theta)/(1-\cos \theta)]$.  
All quantities are measured relative to the $z$ axis in the 
$\gamma^{\ast} \gamma$ c.\ m.\ frame.  
The dominance of the gluon distribution at small $x$ together with 
the LO matrix element for $\gamma^{\ast} g \rightarrow Q \bar{Q}$ 
produces a resolved distribution for $y(c)$ shown in Fig.\ 7(a) 
that is offset from zero.
In contrast, the absence of the gluon distribution in the direct 
case, shown in Fig.\ 7(b), gives a relatively flat distribution.  
The comparison of these two distributions offers a striking illustration of the
effect of the gluon distribution in the photon. Data for these distributions
may well play a significant role in constraining the gluon distribution.

Changes in the above differential distributions with variations in 
the scale are of two types.  First, if the distribution 
is trivial at LO (i.e. delta function like) then the effect of a scale 
variation is an overall shift in the normalization of the curve with little 
effect on the shape.  However, if the distribution is non-trivial at LO, 
then the effect of a scale variation is mild changes in both shape and 
normalization.

%---------------Section 4--------------------------
\section{Conclusion}
%--------------------------------------------------
In this paper we have presented the fully differential heavy quark 
contributions to the photon structure functions of deeply inelastic 
scattering in next-to-leading order QCD.  The calculation was performed 
using the subtraction method.  A variety of distributions 
were discussed for both the resolved and direct components.  Upon 
integration, our fully differential results reduce to, and agree with, 
the single inclusive results already available in the literature.

We studied the renormalization and factorization scale dependence 
of our results.  At the fully inclusive level, the scale variation of 
$F_2^{\gamma}(x,Q^2,m_c^2)$ is greatly reduced at NLO relative to the LO 
calculation.  In addition, the sum of the quark and anti-quark initiated 
process contributions was shown to be small.

The distributions shown in this paper show rather striking differences 
for the direct and resolved cases.  This is due primarily to the presence 
of the gluon distribution in the resolved contribution and lack thereof 
in the direct contribution.   It may therefore be possible to obtain 
information about the gluon distribution in the photon that is complimentary 
to that obtained in other hard scattering processes.

\newpage

%
%----------------------------References-------------------------------------
%

\newpage

\centerline{Figure Captions}

\begin{description}

\item[FIG.1.] 
The process $e^+ + e^- \rightarrow e^+ + e^- + X$.
\item[FIG.2.]
The direct and resolved contributions to $F_2^{\gamma}(x,Q^2,m_c^2)/\alpha$ 
at $Q^2=10\;$(GeV/c)$^2$ as a function of $x$.  The direct contribution 
dominates at large $x$ and the resolved contribution dominates at small $x$.  
The dashed lines are the LO results and the solid lines are the NLO results.
\item[FIG.3.]
The contributions to $d (F_2^{\gamma}(x,Q^2,m_c^2)/ \alpha) / d M_{c \bar{c}}$ 
at $Q^2=10\;$(GeV/c)$^2$.  
a.)\ Predominantly resolved at $x=10^{-3}$ and 
b.)\ predominantly direct at $x=0.3$.  
The dashed lines are the LO results and the solid lines are the NLO results.
\item[FIG.4.]
The contributions to $d (F_2^{\gamma}(x,Q^2,m_c^2)/ \alpha) / d P_t(c)$ at 
$Q^2=10\;$(GeV/c)$^2$.
a.)\ Predominantly resolved at $x=10^{-3}$ and
b.)\ predominantly direct at $x=0.3$.  
The dashed lines are the LO results and the solid lines are the NLO results.
The inset shows the LO direct contribution evaluated at 
$Q^2=100\;$(GeV/c)$^2$ and $x=0.01$.
\item[FIG.5.]
The contributions to 
$d (F_2^{\gamma}(x,Q^2,m_c^2)/ \alpha) / d P_t(c \bar{c})$ at 
$Q^2=10\;$(GeV/c)$^2$.
a.)\ Predominantly resolved at $x=10^{-3}$ and
b.)\ predominantly direct at $x=0.3$.
The dashed lines are the LO results and the solid lines are the NLO results.  
\item[FIG.6.]
The contributions to
$d (F_2^{\gamma}(x,Q^2,m_c^2)/ \alpha) / d (\Delta \phi)$ at
$Q^2=10\;$(GeV/c)$^2$.
a.)\ Predominantly resolved at $x=10^{-3}$ and
b.)\ predominantly direct at $x=0.3$.
The dashed lines are the LO results and the solid lines are the NLO results.
\item[FIG.7.]
The contributions to $d (F_2^{\gamma}(x,Q^2,m_c^2)/ \alpha) / d y (c)$ at
$Q^2=10\;$(GeV/c)$^2$.
a.)\ Predominantly resolved at $x=10^{-3}$ and
b.)\ predominantly direct at $x=0.3$.
The dashed lines are the LO results and the solid lines are the NLO results.
\item[FIG.8.]
The contributions to $d (F_2^{\gamma}(x,Q^2,m_c^2)/ \alpha) / d \eta (c)$ at
$Q^2=10\;$(GeV/c)$^2$.
a.)\ Predominantly resolved at $x=10^{-3}$ and
b.)\ predominantly direct at $x=0.3$.
The dashed lines are the LO results and the solid lines are the NLO results.

\end{description}

\newpage
\pagestyle{empty}
\begin{figure}
\centerline{\hbox{\psfig{figure=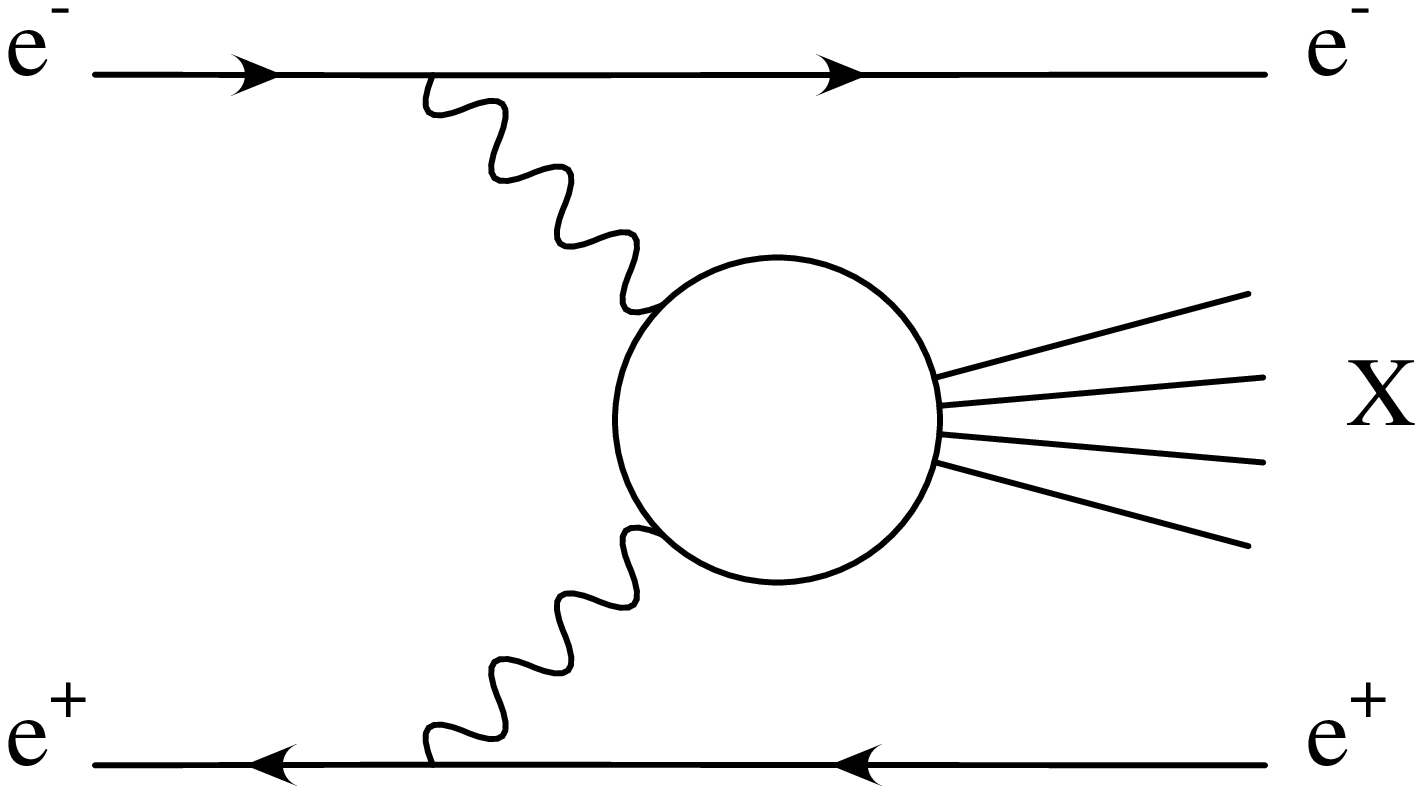}}}
\end{figure}
\newpage

\begin{figure}
\centerline{\hbox{\psfig{figure=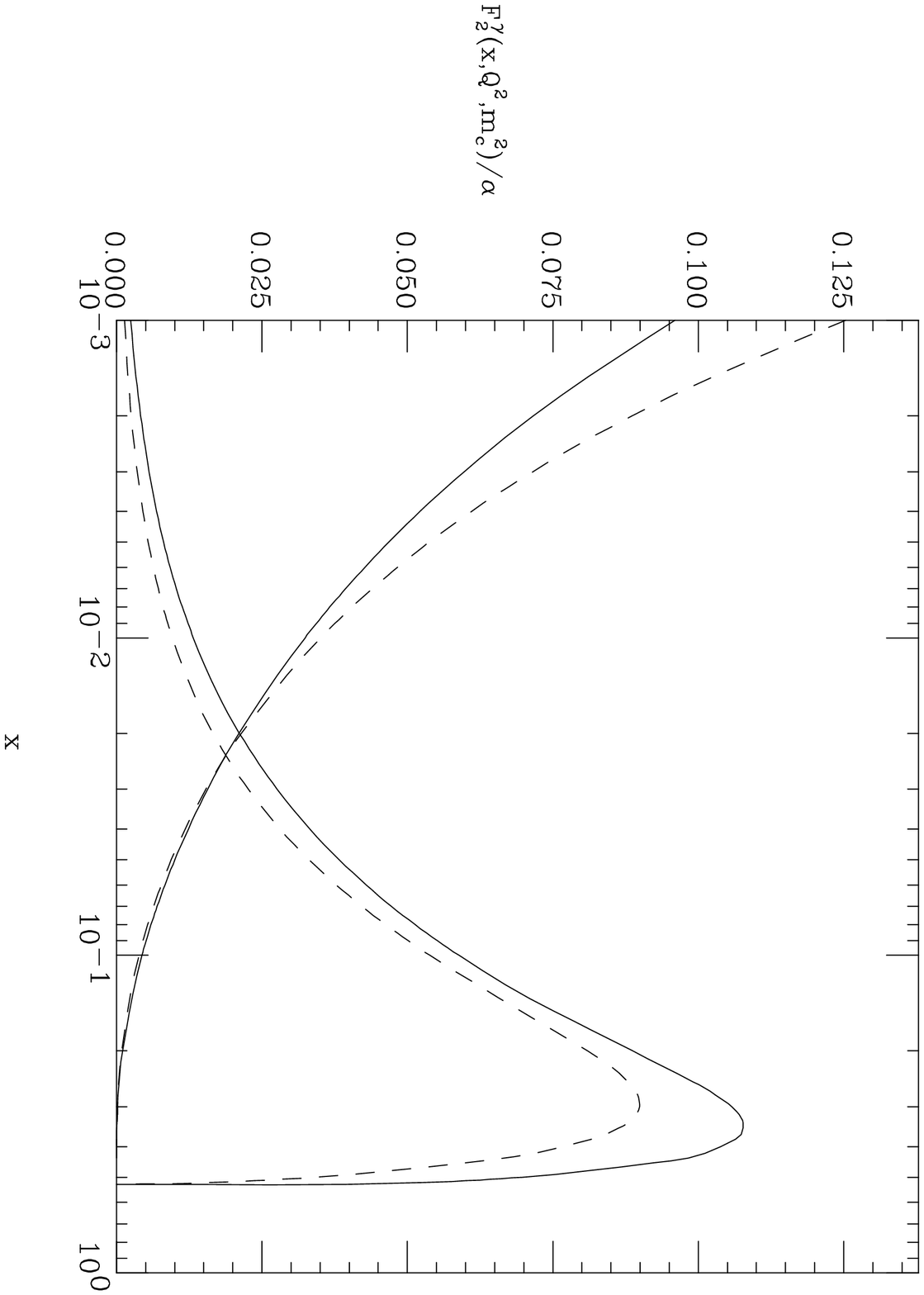}}}
\end{figure}
\newpage

\begin{figure}
\centerline{\hbox{\psfig{figure=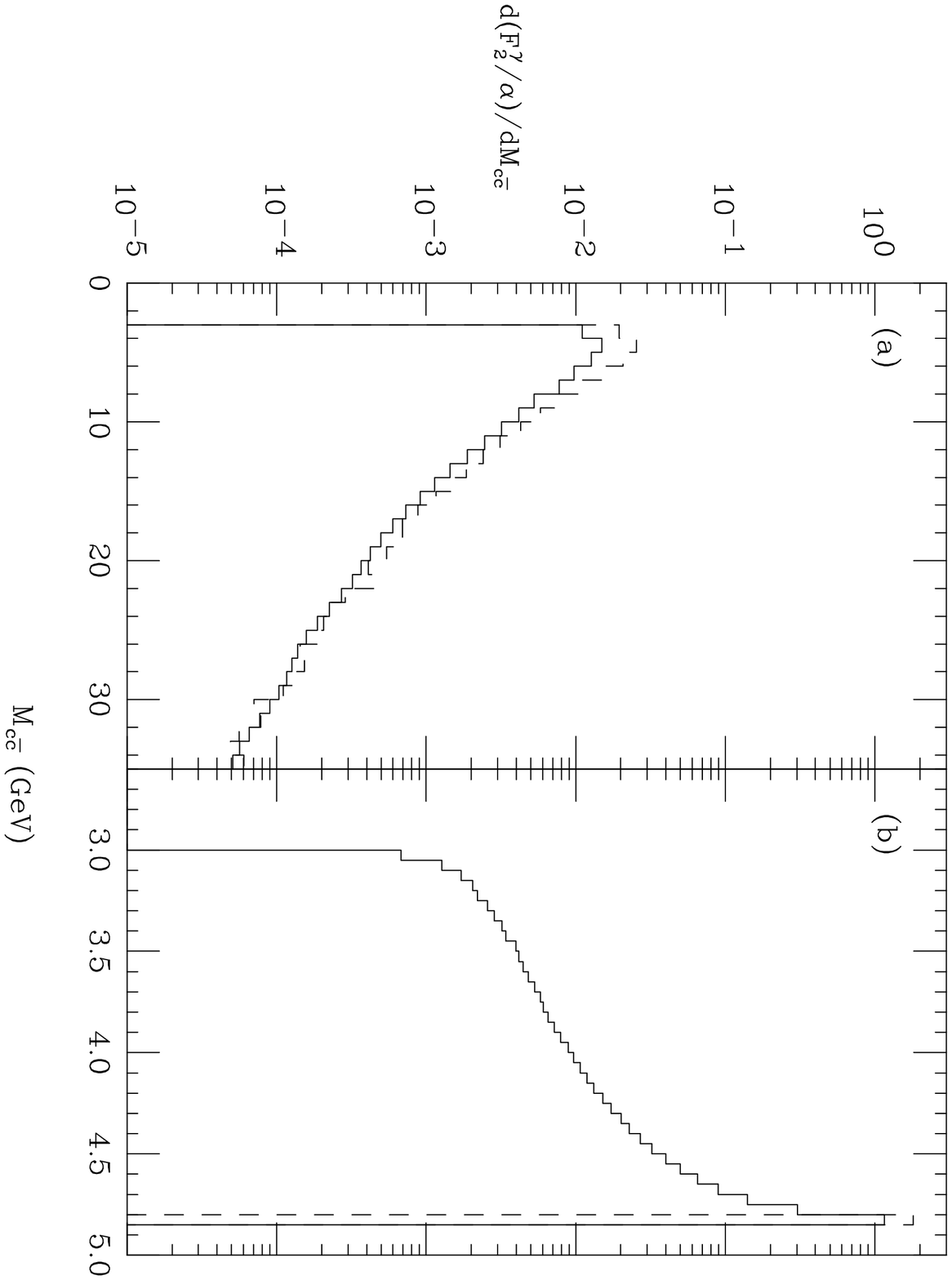}}}
\end{figure}
\newpage

\begin{figure}
\centerline{\hbox{\psfig{figure=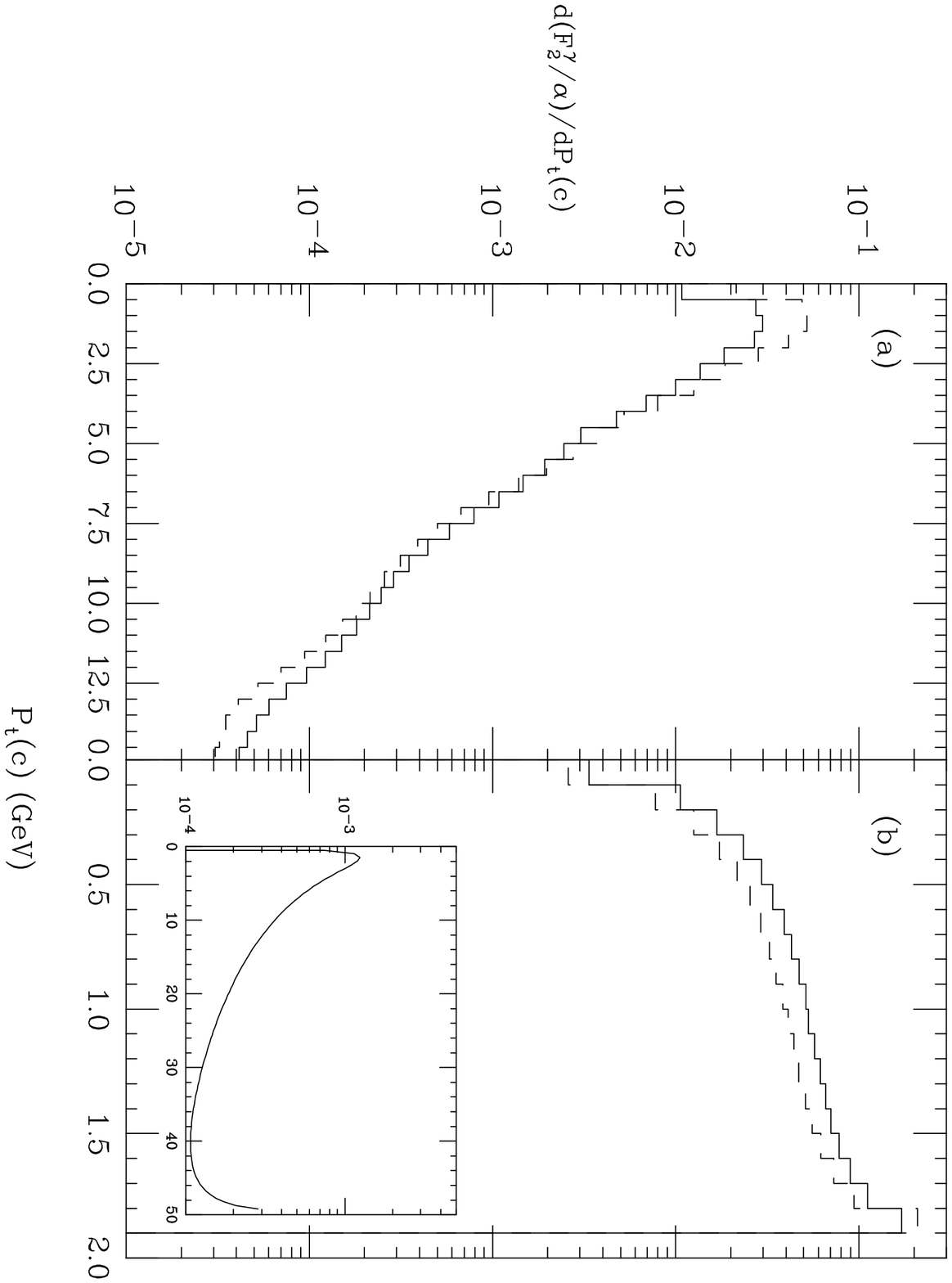}}}
\end{figure}
\newpage

\begin{figure}
\centerline{\hbox{\psfig{figure=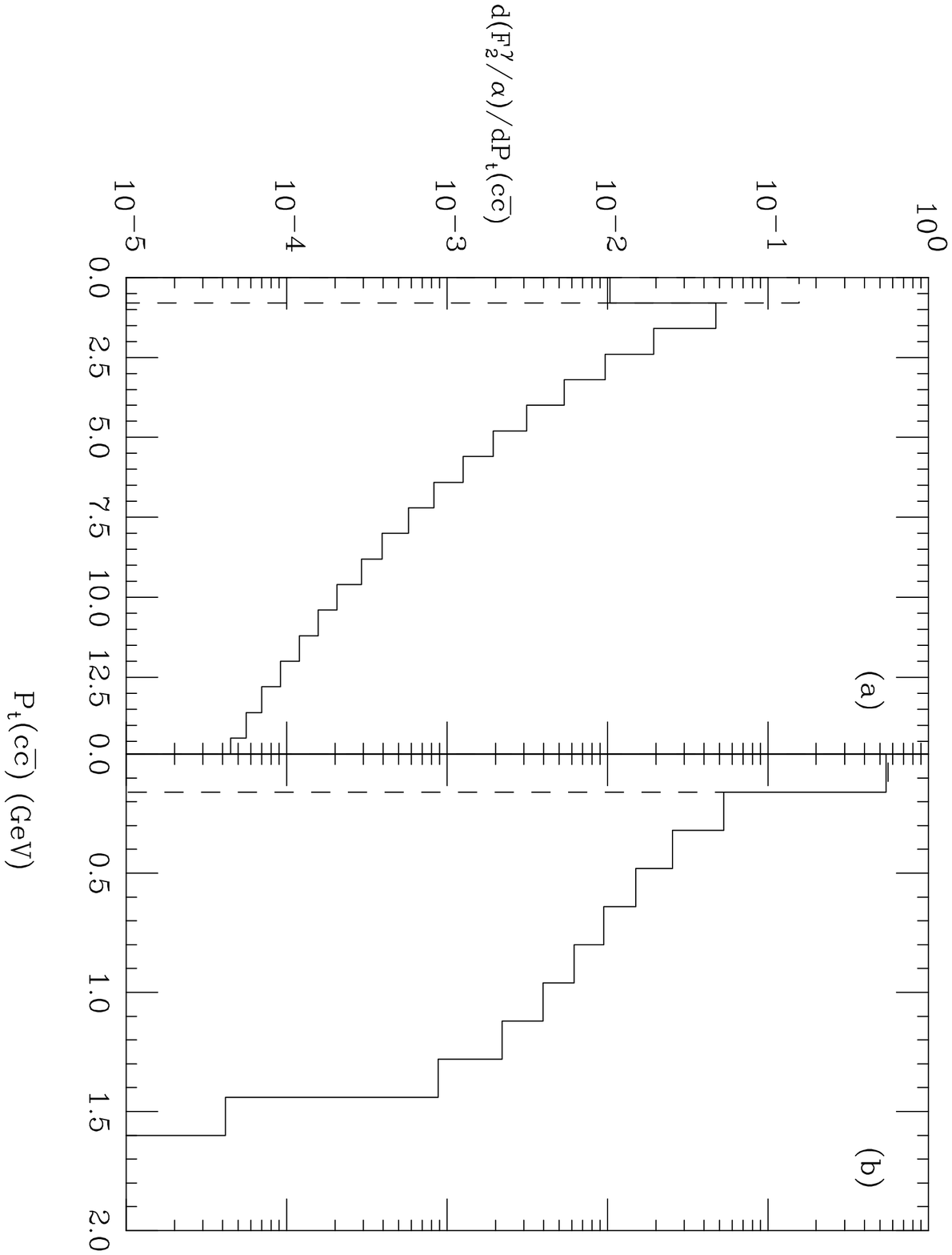}}}
\end{figure}
\newpage

\begin{figure}
\centerline{\hbox{\psfig{figure=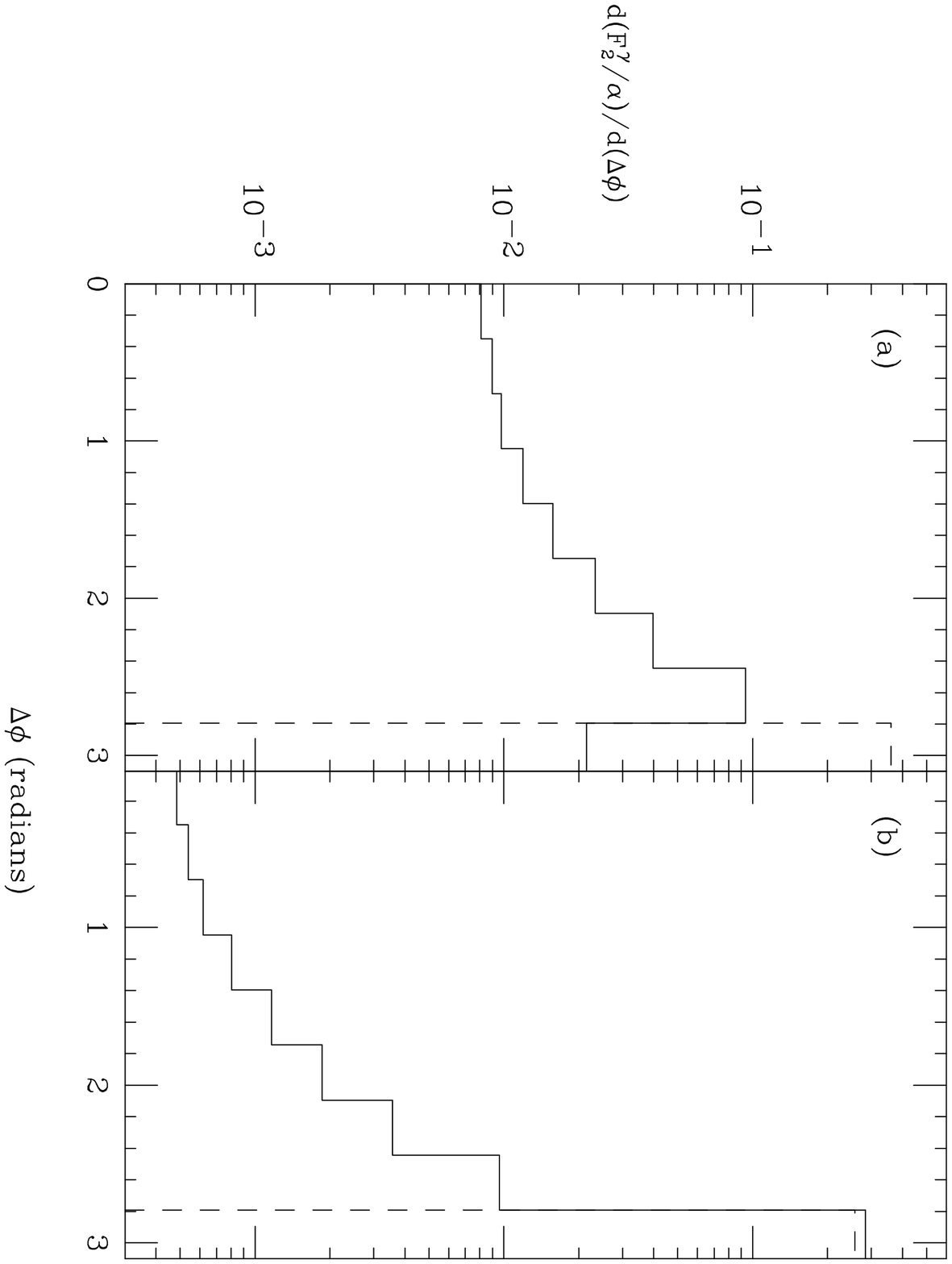}}}
\end{figure}
\newpage

\begin{figure}
\centerline{\hbox{\psfig{figure=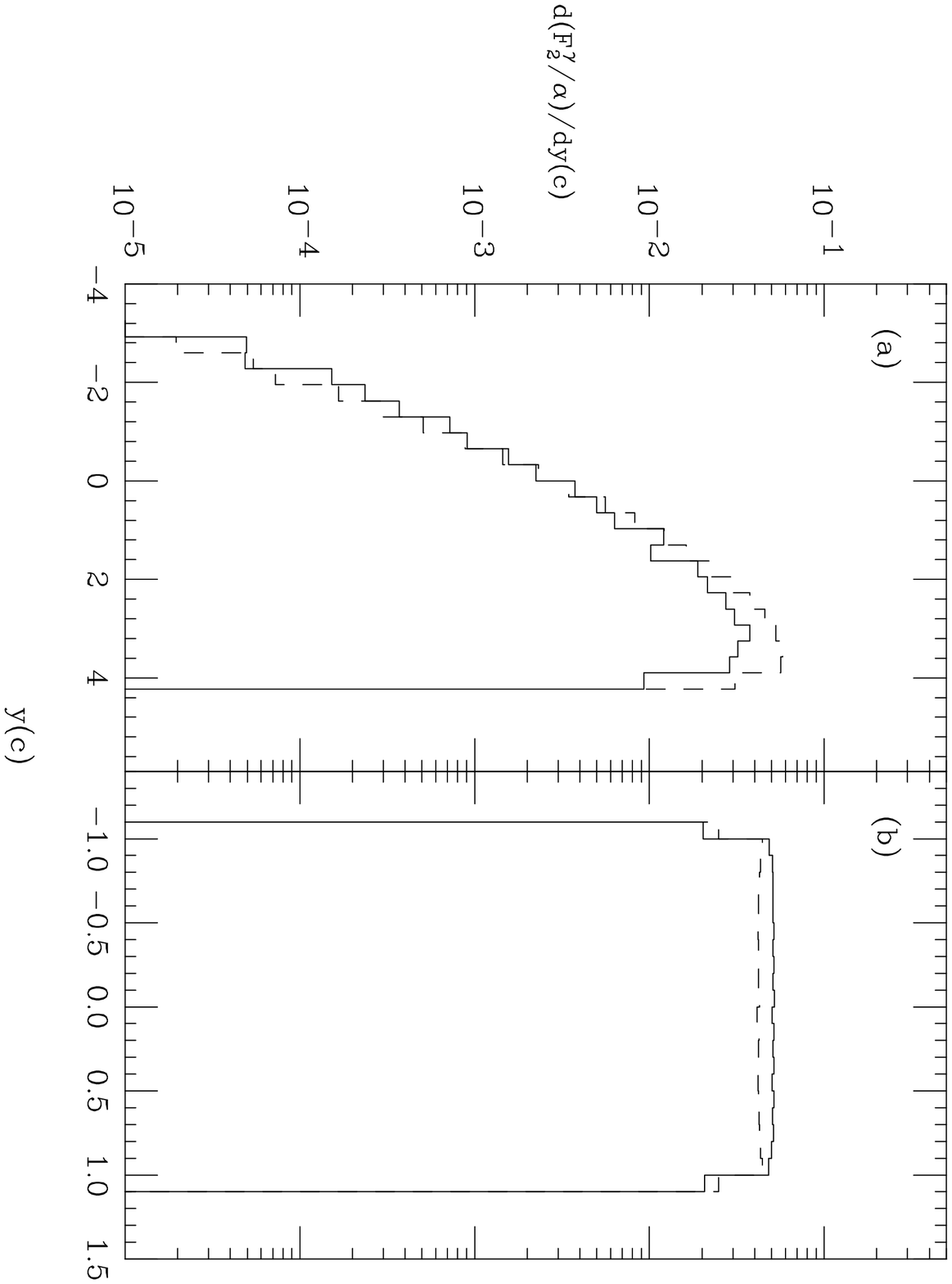}}}
\end{figure}
\newpage

\begin{figure}
\centerline{\hbox{\psfig{figure=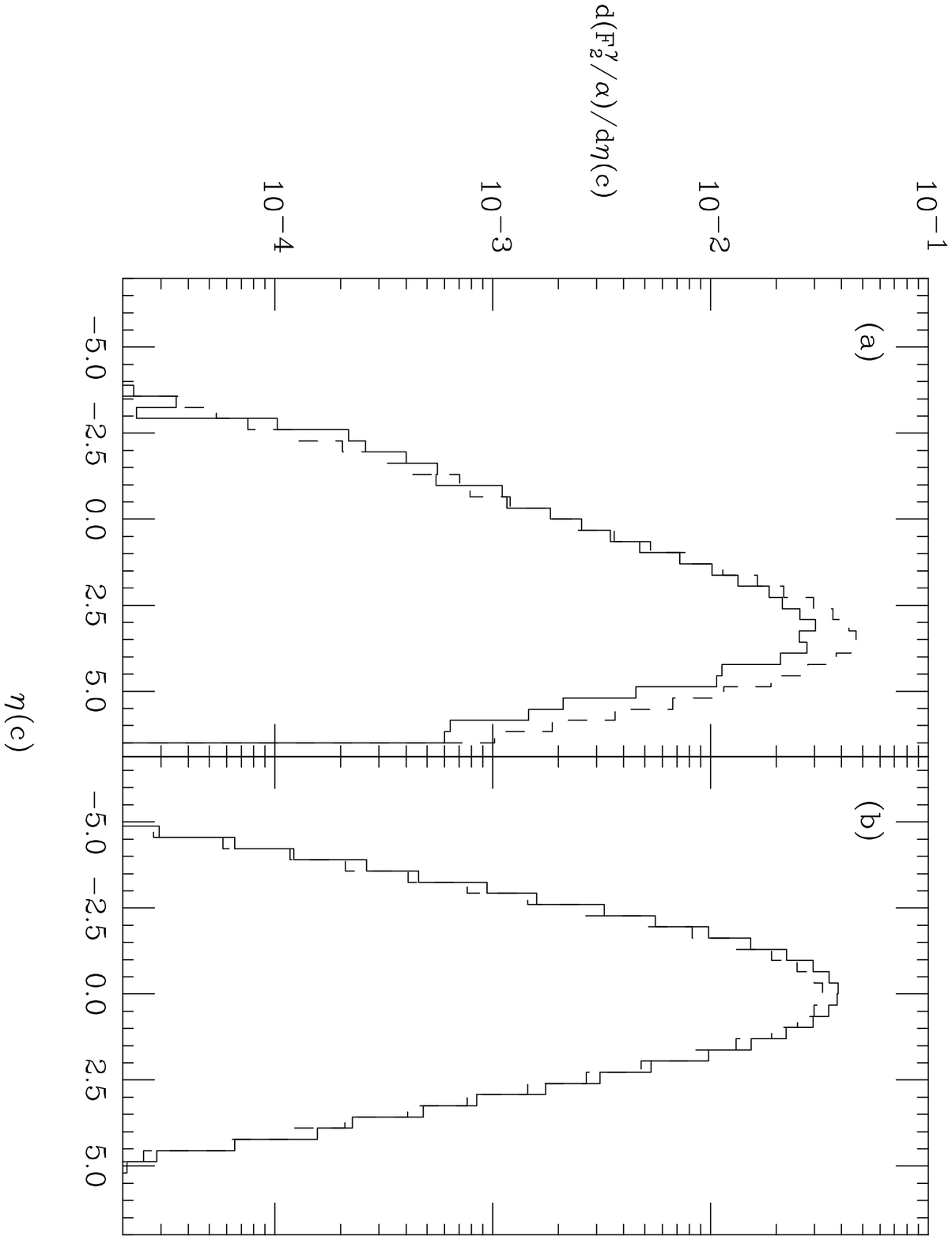}}}
\end{figure}
\newpage

\end{document}